# Structural and ferroelectric properties of perovskite $Pb_{(1-x)}(K_{0.5}Sm_{0.5})_xTiO_3$ ceramics

Arun Kumar Yadav[1], Anita Verma[1], Sunil Kumar[1], Anjali panchwanee[3], V. Raghavendra Reddy[3], Parasharam M. Shirage[1,2], Sajal Biring[4], Somaditya Sen[1,2]

[1]Discipline of Metallurgy Engineering and Materials Science, Indian Institute of Technology, Khandwa Road, Indore-453552, India
[2]Discipline of Physics, Indian Institute of Technology, Khandwa Road, Indore-453552, India
[3]UGC-DAE Consortium for Scientific Research, University Campus, Khandwa Road, Indore-452001, India
[4]Electronic Engg., Ming Chi University of Technology, New Taipei City, Taiwan

Corresponding Author: sens@iiti.ac.in

**Abstract:**

$PbTiO_3$ has the highest tetragonal distortion ($c/a \sim 1.064$) and highest spontaneous polarization among perovskite titanates. But, it is hazardous and hence one needs to reduce $Pb$ content by substituting or reducing $Pb$ content for use in applications. $Pb_{(1-x)}(K_{0.5}Sm_{0.5})_xTiO_3$ ($0 \leq x \leq 0.5$) perovskite powders were synthesized by sol-gel process, where $Pb^{2+}$ was replaced by a combination of $K^{+1}_{0.5}Sm^{+3}_{0.5}$ (equivalent charge and comparable ionic radius) providing an excellent substitution model to study changes in structural and electrical properties. Vibrational properties and dielectric properties are modified with substitution. A polar *tetragonal* to a nearly nonpolar *cubic* phase transition decreases to lower temperatures with substitution due to reduces the lattice strain with substitution. Ferroelectricity is retained even for $x=0.5$, which has a nearly *cubic* phase and makes the material technologically important.

## 1. Introduction

Properties like ferroelectricity, piezoelectricity, and pyroelectricity were discovered in perovskite $ABO_3$ compounds. Ferroelectrics are used in energy conversion devices, ultrasonic medical diagnostic apparatus, ultrasonic non-destructive detectors, pyroelectric infrared sensors and magnetoelectric sensors [1-3]. $PbTiO_3$ is a $ABO_3$ perovskite, a dominant ferroelectric material used in sensors, actuators, capacitors, nonvolatile memories, ultrasonic transducers, high piezoelectric, electromechanical and electro-optic properties etc.[4-5]. $PbTiO_3$ has a structural phase transition from tetragonal ($P4mm$) to cubic ($Pm3m$) at 763K. It has a highly strained lattice due to strong orbital hybridization and subsequent non-centrosymmetric lattice distortion. Among the perovskites, $PbTiO_3$ has the highest tetragonal distortion ($c/a \sim 1.064$) which makes it significantly more ferroelectric with highest spontaneous polarization [6-7]. But, *PbTiO$_3$* not being an environmental friendly material due to high toxic nature of *Pb,* there is a necessity to reduce *Pb* content in applications from safety and health concerns. Substituted *PbTiO$_3$* compounds have been explored to investigate the science and applicability of these materials. A few such ferroelectric ceramics materials are *Pb(Mg$_{1/3}$Nb$_{2/3}$)O$_3$-xPbTiO$_3$, Ba(Mg$_{1/3}$Nb$_{2/3}$)O$_3$−xPbTiO$_3$, Ba(Zn$_{1/3}$Nb$_{2/3}$)O$_3$−xPbTiO$_3$, Ba(Yb$_{1/2}$Nb$_{1/2}$)O$_3$−xPbTiO$_3$, Ba(Sc$_{1/2}$Nb$_{1/2}$)O$_3$−xPbTiO$_3$, BaSnO$_3$−xPbTiO$_3$, (1−x)PbTiO$_3$−xBi(Zn$_{1/2}$Ti$_{1/2}$)O$_3$, (1−x)PbTiO$_3$−xBi(Ni$_{1/2}$Ti$_{1/2}$)O$_3$ and (1− x)PbTiO3–xBiFeO3* etc. [8-13]. By making isovalent or hetero-valent substitution on *Pb$^{2+}$* site, lattice anisotropy is reduced. A polar *tetragonal* to nonpolar *cubic* phase transition decreases to lower temperatures mostly decreases with substitution as most substitution reduces the lattice strain. Also the phase transition gets diffused due to cumulative structural transition resulting from the different *T$_C$* for localized region in given system, resulting in a broad peak with temperature in dielectric response rather than sharp peak in normal ferroelectric [14-16].

[1]

A combination of *K$^{+1}_{0.5}$Sm$^{+3}_{0.5}$* provides the same charge and a comparable ionic radius as compared to *Pb$^{2+}$*. Thus this combination may provide an excellent substitution model enabling a detailed study of changes in the physical properties. In this study we report for the first time a new series, *Pb$_{(1-x)}$(K$_{0.5}$Sm$_{0.5}$)xTiO$_3$* (0≤*x*≤*0*.50). Detailed studies of the structure, vibrational, dielectric and ferroelectric properties are analyzed.

## 2. Materials and methods

Sol-gel processed polycrystalline *Pb$_{(1-x)}$(K$_{0.5}$Sm$_{0.5}$)$_x$TiO$_3$* (0≤*x*≤0.5) (*PKST*) ceramics were prepared using hydrated Lead (II) nitrate (*Pb(NO$_3$)$_2$*), Potassium nitrate (*KNO$_3$*), Samarium oxide and Titanium(IV) bis(ammonium lactato) dihydroxide (*TALH*) solution as precursors with purity>99.999% (Alfa Aesar, Puratronic grade). *Sm$_2$O$_3$* was dissolved in dilute *HNO$_3$*. *TALH*, *Pb(NO$_3$)$_2$* and *KNO$_3$* were dissolved in double distilled water. *Sm* and *Ti* solutions were mixed together and thereafter the *K* solution was added to form a mixed solution of *Sm, Ti* and *K* and stirred for a while. The *Pb* solution was added to this combined solution. A solution of citric acid and ethylene glycol of 1:1 molar ratio was prepared in a separate beaker as gel former and was thereafter added to the mixture. Appropriate amount of ammonium hydroxide was added to the solution to maintain a pH=8.5. The resultant sols were vigorously stirred and heated at ~85$^o$C on a hot plate to form gels. The gels were burnt to form black powders. To de-carbonate and de-nitrate, these powders were heated at 500$^o$C for 12h.

Pellets (diameter~10mm; thickness~1.5mm) were prepared by uni-axially pressing the carefully grinded powders mixed with a binder *PVA* solution. The pellets were sintered at 600$^o$C for 6h to burn the binder followed by another heating at 1050$^o$C for 6h to form mechanically dense pellets. Densities of the samples were estimated from the weights and dimensions of the pellets. The theoretical densities were also estimated using formula weight and volume of unit cells calculated from the refined lattice parameters of the samples. Relative densities range from

~ 93%, 94%, 95%, 95%, 95%, 92%, and 91% for compositions with $x$=0.06, 0.09, 0.12, 0.18, 0.25, 0.37, and 0.50, respectively.

X-ray diffraction (XRD) was performed using Bruker D2 Phaser x-ray Diffractometer to ensure the phase formation and purity of the sintered samples. Raman spectroscopy was performed using a Czerny-Turner type achromatic spectrograph with spectral resolution of 0.4 cm$^{-1}$/pixel and excitation of 632.8nm. Microstructure and the grain size analysis of the sintered pellets were investigated by Supra55 Zeiss field emission scanning electron microscope. Electrodes were prepared using silver paste painted on both sides of the sintered pellet. Silver coated pellets were cured at 550$^o$C for 10 minutes. Before doing the measurement, we heated the samples at 200$^o$C for 10 minutes to avoid any moisture content.

A Newtons 4$^{th}$ LTD PSM 1735 LCR meter was used to study the dielectric properties. Ferroelectric (P-E) studies were carried out by ferroelectric loop (P-E) tracer (M/s Radiant Instruments, USA). During the ferroelectric measurements, the samples were immersed in silicone oil to prevent electric arcing, at high applied voltages.

## 3. Results and discussion
### 3.1 Structural properties

X-ray powder diffraction (XRD) patterns of the PKST powders with 0≤$x$≤0.5, were analyzed after crushing the pellets sintered at 1050$^o$C. The XRD patterns are shown in Figure 1a. With increasing substitution, the separation between the <001> and <100> peaks reduces. All the compositions belong to a tetragonal space group *P4mm* phase. A very weak PbTiO$_3$ *I4/m* space group contribution is also observed [Figure 1a]. A Le Bail profile fitting of the XRD data was done with Topas3.2 software to estimate the lattice parameters. The goodness of the fitting for $x$=0.06 is shown in Figure 1d. Lattice parameter '*c*' decreases with increasing composition *x*, while '*a*' is almost constant (Figure 1b). Hence the *c/a* ratio decreases to ~1 for $x$=0.5 (Figure 1c). The variation of lattice parameters with composition is also a signature of proper substitution of *Pb* by *K/Sm*.

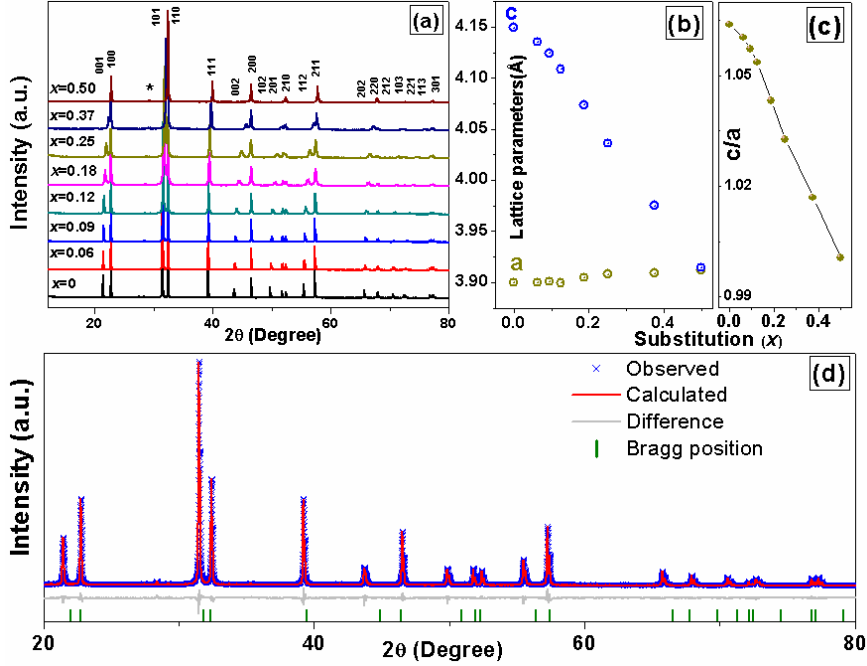

**Figure 1:** (a) X-ray powder diffraction patterns of $Pb_{(1-x)}(K_{0.5}Sm_{0.5})xTiO_3$ samples, (b) Lattice parameters, (c) tetragonal strain, $c/a$ ratio at room temperature. (d) Le Bail profile fitting of $Pb_{(1-x)}(K_{0.5}Sm_{0.5})xTiO_3$ for $x=0.06$ composition where the crosses are observed data points, the solid line is calculated data and line below is the difference between observed and calculated data. The vertical bars are the Bragg position of the reflection in the *P4mm* space group.

In perovskite $ABO_3$ samples, tolerance factor is a quantitative measure of the mismatch between the bonding requirements of *A* and *B*-site cations and subsequently reflects structural distortions such as rotation/tilt of the *B*-octahedral. Lattice stability and distortion of crystal structures are related to this mismatch [17] and hence can be estimated from the tolerance factor (*t*), given by,

$$t = \frac{(R_A+R_O)}{\sqrt{2}(R_B+R_O)} \qquad (1)$$

where, $R_A$, $R_B$ and $R_O$ are the radii of *A* and *B*-site ions and *O* ions, respectively. Since the average *A*-site ionic radii of the substituent are lesser compared to *Pb*, tilt as well as non-centrosymmetric distortion will reduce. The merging peaks in the XRD data indicate a transformation from non-centrosymmetric to centrosymmetric structure [18-19]. The tolerance factor of the PKST samples decreases from 1.019 as in $x=0$ to 0.988 for $x=0.5$, also the strain decreases with increasing substitution.

Room temperature Raman spectroscopy [Figure 2] was done to qualitatively assess retention of domain structure, defects and structural distortions and thereby understand deformations and lattice strains associated with substitution in PKST. All the observed modes belongs to pure *PbTiO₃* having tetragonal space group symmetry ($C_{4v}^1$) with *12* optical modes attributed to three $A_1$-symmetry modes, eight *E*-symmetry modes (Four degenerate pairs) and a $B_1$-symmetry mode [21-22].

[4]

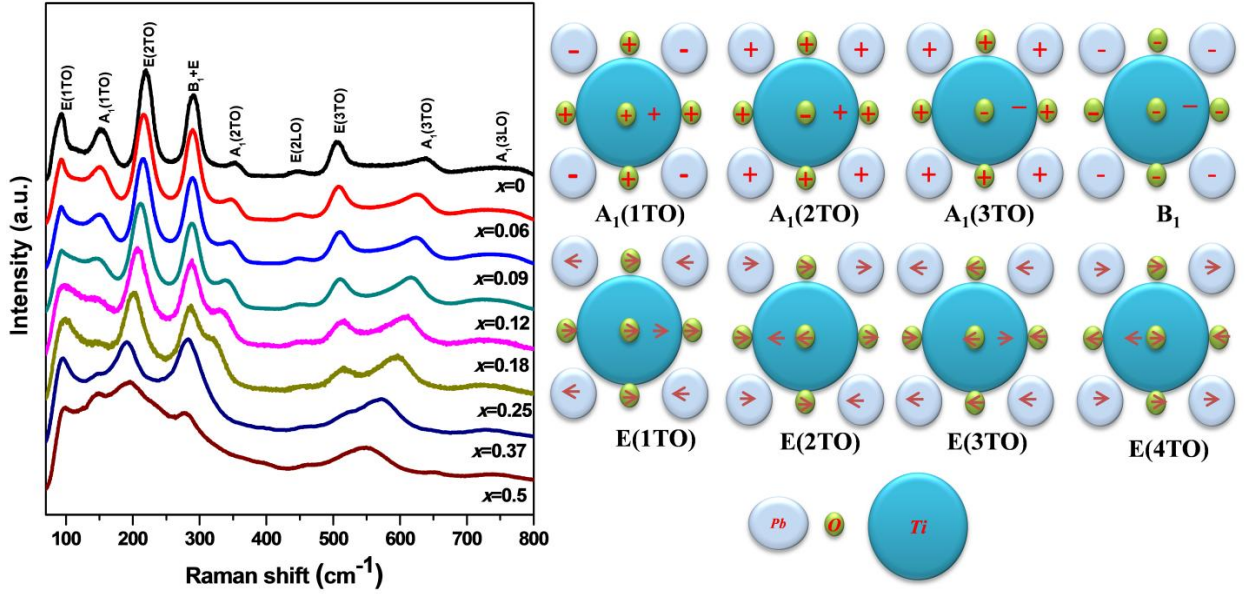

**Figure 2:** (Left) Raman spectra of $Pb_{(1-x)}(K_{0.5}Sm_{0.5})xTiO_3$ samples at room temperature. (Right) Cartoon shows different modes and their related vibration type [20].

Intensity of the *$A_1(1TO)$*, *$A_1(2TO)$* and *$A_1(3TO)$* reduces with increasing substitution. Among the *$A_1$* modes the most intense mode is *$A_1(1TO)$*. This mode merges with the *$E(1TO)$* mode to form the *$T_{1u}$* mode in the cubic form. The *$A_1(1TO)$* mode is due to oscillations of the *Ti*-octahedra (i.e. *Ti* and *O* together) relative to *Pb* ions [23]. This mode does not soften much but nearly vanishes after *x*=0.09 and reappears again for 0.25≤*x*≤0.5. The $A_1(2TO)$ mode is due to oscillations of the *Ti* ion relative to *O* and *Pb* ions. This mode softens nominally until *x*=0.09 but thereafter drastically softens to merge with the *$E(2TO)$* mode in *x*=0.5. A similar but less drastic trend is observed for the *$A_1(3TO)$* mode, in which the *Ti* ions and the apical *O* ions together form a chain and vibrates along the *c*-axis with respect to the *Pb* and the other *O* atoms. From all these three modes it is understood that relative motion between the *Pb, Ti* and *O* ions reduces along the *c*-axis with substitution. The tetragonal *$PbTiO_3$* is a strained lattice. The strain originates from a very strong hybridization between *$Pb(6s^2)$-$O(2p)$* and also due to a moderate *Ti*(3d)-*O*(2p) hybridization [2, 24]. As a result of substitution of *Pb* by *K/Sm*, we expect a reduction of the strong *$Pb(6s^2)$-$O(2p)$* bond. This reduction in the strength of the bond reduces energy of vibration which is reflected in the softening of the $A_1(2TO)$. Such a reduction in energy in the *$A_1(1TO)$* mode is not observed as in this mode, the A-site is freely oscillating with respect to the *Ti-O* octahedron. However the lesser number of oscillators reduce the intensity of $A_1(1TO)$ and *$A_1(2TO)$*. Note that the *$E(3TO)$* mode gains energy with the substitution. The lower mass of *K/Sm* compared to *Pb* may be a reason for the same.

The most noticeable significance of the Raman measurement is that even at *x*=0.50 we observe all the modes where from XRD *c/a* ~1. Such an observation hints at the fact that a remnant strain which may or may not be related to the tetragonal structure is retained in the high

[5]

$x$ samples. In the cubic mode, pure *PbTiO₃* does not have any Raman mode, as all the *A₁(TO)* and *E(TO)* modes merge and vanish gradually at the phase transition [20].

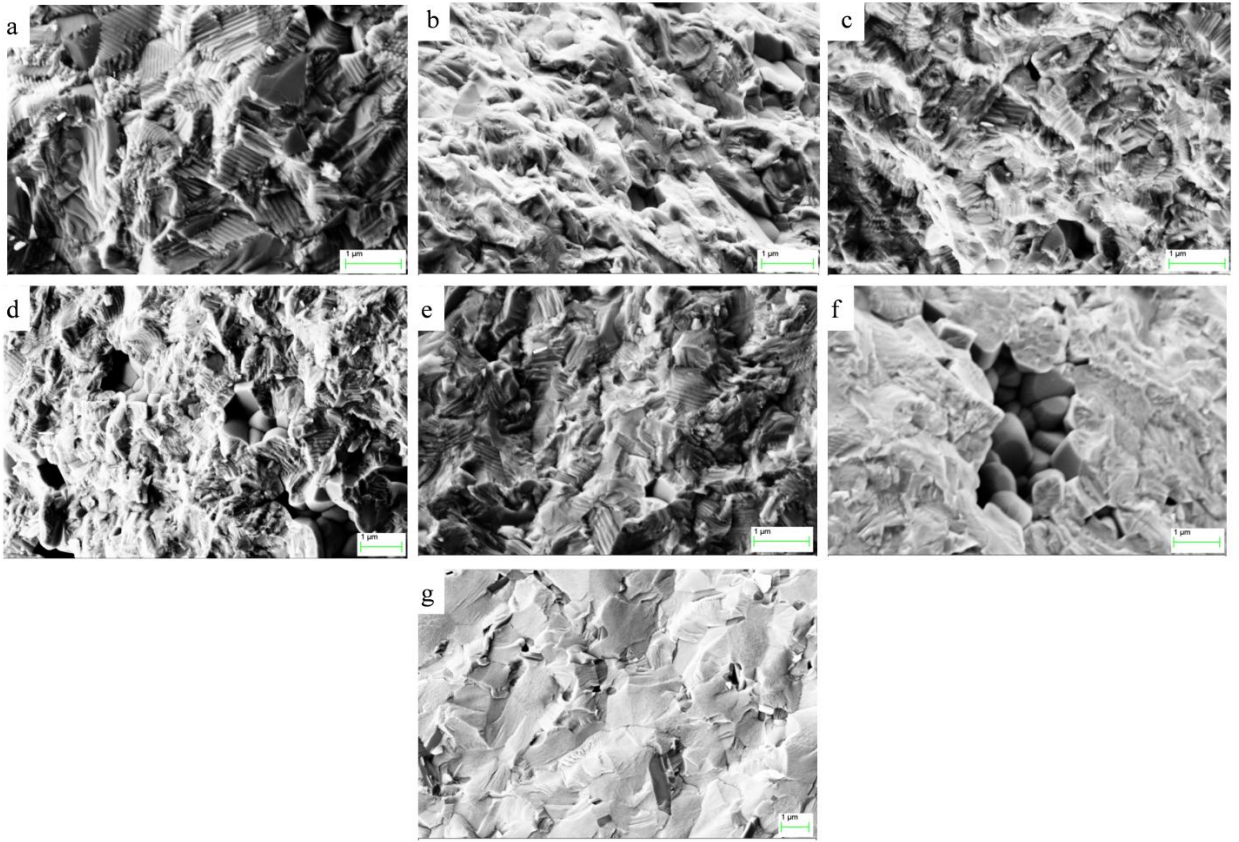

**Figure 3**. Microstructure analysis of fractured surface of Pb$_{(1-x)}$(K$_{0.5}$Sm$_{0.5}$)$_x$TiO$_3$ samples, (a) $x=0.06$, (b) $x=0.09$, (c) $x=0.12$, (d) $x=0.18$, (e) $x=0.25$, (f) $x=0.37$, and (g) $x=0.50$ compositions.

Fractured surface FESEM images of PKST ceramic samples are recorded in $0.06 \leq x \leq 0.50$ range (Figure 3). Grains are closely packed in all the samples. The average grain size calculated from the FESEM is $1.19 \pm 0.53$ μm for the composition with $x=0.06$ which decreases to $0.98 \pm 0.37$ μm for $x=0.50$. The average grain size seams to decrease nominally with increasing *K/Sm* substitution. This behavior could be attributed to lower diffusivity of rare earth during sintering. Similar behavior has been observed in other rare earth doped perovskites and layered perovskites [25-26].

### 3.2 Dielectric properties

Room temperature (RT) dielectric properties were measured as a function of frequency for PKST ($0.06 \leq x \leq 0.50$). As shown in Figure 4a and Figure 1S, with increasing frequency the value of dielectric constant ($\varepsilon'$) and loss (tan$\delta$) decrease at low frequency range (100 Hz to 10 kHz) and remains almost constant in high frequency region. Space charge polarization may contribute to higher values of dielectric constant. We observe that with increasing substitution, the value of $\varepsilon'$ of PKST ceramics increases exponentially [Figure 4b]. This behavior of $\varepsilon'$ at room temperature and similar trend in *tanδ* can be attributed to the higher dipolar polarizability of *K/Sm* than *Pb* [27-28]. For *x*=0.50, $\varepsilon'$ increases rapidly probably due to its near room temperature phase transition.

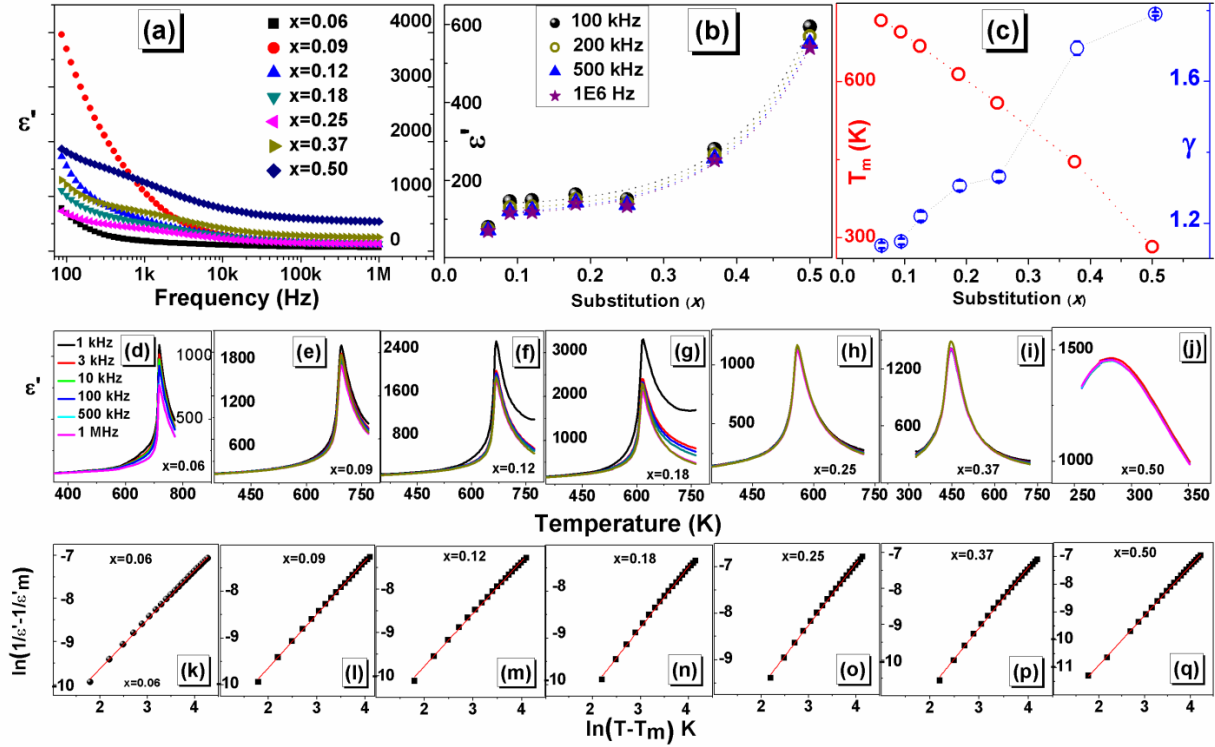

**Figure 4:** (a) Frequency dependence of dielectric constant of $Pb_{(1-x)}(K_{0.5}Sm_{0.5})_xTiO_3$, (b) composition dependence of high frequency dielectric constant at room temperature;(c) Phase transition temperature ($T_m$) and degree of diffusivity ($\gamma$) with composition $x$; (d-j) Permittivity data for high frequency as a function of temperature for PKST samples; (k-q) linear fitting of $\ln[(1/\varepsilon')-(1/\varepsilon'_m)]$ verses $\ln(T-T_m)$ (modified Curie-Weiss law) at 1MHz, solid lines representing the fitted data.

To study the phase transition in PKST and other dielectric properties, capacitance and dielectric loss factor were measured over the temperature range of 300K – 773K at various frequencies in 1 kHz –1 MHz range. It was quite difficult to fabricate a pure phase, sufficiently dense and mechanically robust *PbTiO₃* pellet. Hence, dielectric properties of *PbTiO₃* phase were not investigated.

Pure *PbTiO₃* undergoes a sharp ferroelectric to paraelectric transition accompanied by a structural transition from polar tetragonal to non-polar cubic phase [29]. In such ferroelectrics, we observe maxima in the $\varepsilon'$-$T$ data, associated with this phase transition. For *PbTiO₃* the reported phase transition temperature is around 763 K (Curie temperature). In the PKST samples this phase transition temperature ($T_m$) varies from 680K for *x*=0.06 to 341 K as in *x*=0.50. Note that the transition temperature is not frequency dependent; thereby the samples are not relaxors.

The phase transition temperature approximately decreases linearly with substitution. [Figure 4c] The *Pb 6s²* lone pair helps in stabilizing the tetragonal strain (~ 6%) in *PbTiO₃* [30]. We have shown from our XRD and Raman analysis that with increasing substitutions the tetragonal strain is relieved in PKST. The reduction of non-centrosymmetric distortion reduces with substitution thereby requiring lesser thermal energy to achieve centrosymmetric cubic structure. Therefore with increasing *x*, phase transition temperature decreases. Similar trend has been observed in other perovskite related ferroelectrics [31].

[7]

Another feature in [Figure 4d to 4j] is an apparent increase in diffuseness of peak in $\varepsilon'$-$T$ plots. Diffuse phase transition is usually observed in perovskites with random distribution of different types of ions on structurally identical sites in lattice. It must be noted that diffuse phase transition exhibit a broad change of structure and properties at the Curie point compared to a sharp peak in normal ferroelectric materials [32-34]; consequently, the phase transition characteristics of such materials are known to diverge from the characteristic of Curie-Weiss behavior and can be described by a modified Curie-Weiss formula [35-36].

$$\frac{1}{\varepsilon'} - \frac{1}{\varepsilon'_m} = C^{-1}(T - T_m)^\gamma \qquad (2)$$

Where, *C* is Curie-Weiss constant and $\gamma$ (1≤$\gamma$≤2) gives the degree of diffuseness. A value of $\gamma = 2$ describes an ideal diffuse phase transition. The degree of diffuseness was calculated by the least-square linear fitting of ln $(\frac{1}{\varepsilon'} - \frac{1}{\varepsilon'_m})$ versus ln (T-$T_m$) curves at a frequency of 1 MHz of PKST ceramics. The slope of linear fit, was found to increase from 1.136±0.007 for *x*=0.06 to 1.693±0.019 for *x*=0.50 indicating a significant increase in diffuseness of phase transition in doped samples (Figure 4c). Compositional disorder arising due to the random distribution of $K^+$ and $Sm^{3+}$ seems to be responsible for observed diffuse phase transition in PKST ceramics.

Impedance spectroscopy enables to determine the various contributions to dielectric constant, such as grain, grain boundaries and electrode effect [37]. Impedance of such type of system can be described by the Cole-Cole equation [38]

$$Z^*(\omega) = \frac{R_1}{1+(j\omega\tau)^{1-\alpha}} \qquad (3)$$

The real (Z') and imaginary (Z") parts of impedance can be written in the following manner:

$$Z'(\omega) = \frac{R_1[1+(\omega\tau)^{1-\alpha}\sin(\frac{\alpha\pi}{2})]}{1+2(\omega\tau)^{1-\alpha}\sin(\frac{\alpha\pi}{2})+(\omega\tau)^{2-2\alpha}} \qquad (4)$$

$$Z"(\omega) = \frac{R_1(\omega\tau)^{1-\alpha}\cos(\frac{\alpha\pi}{2})}{1+2(\omega\tau)^{1-\alpha}\sin(\frac{\alpha\pi}{2})+(\omega\tau)^{2-2\alpha}} \qquad (5)$$

with Cole-Cole parameter, $\alpha=2\theta/\pi$ being a measure of distribution of relaxation time, $\tau=RC$, related to the resistance *R* and capacitance *C* associated with grain.

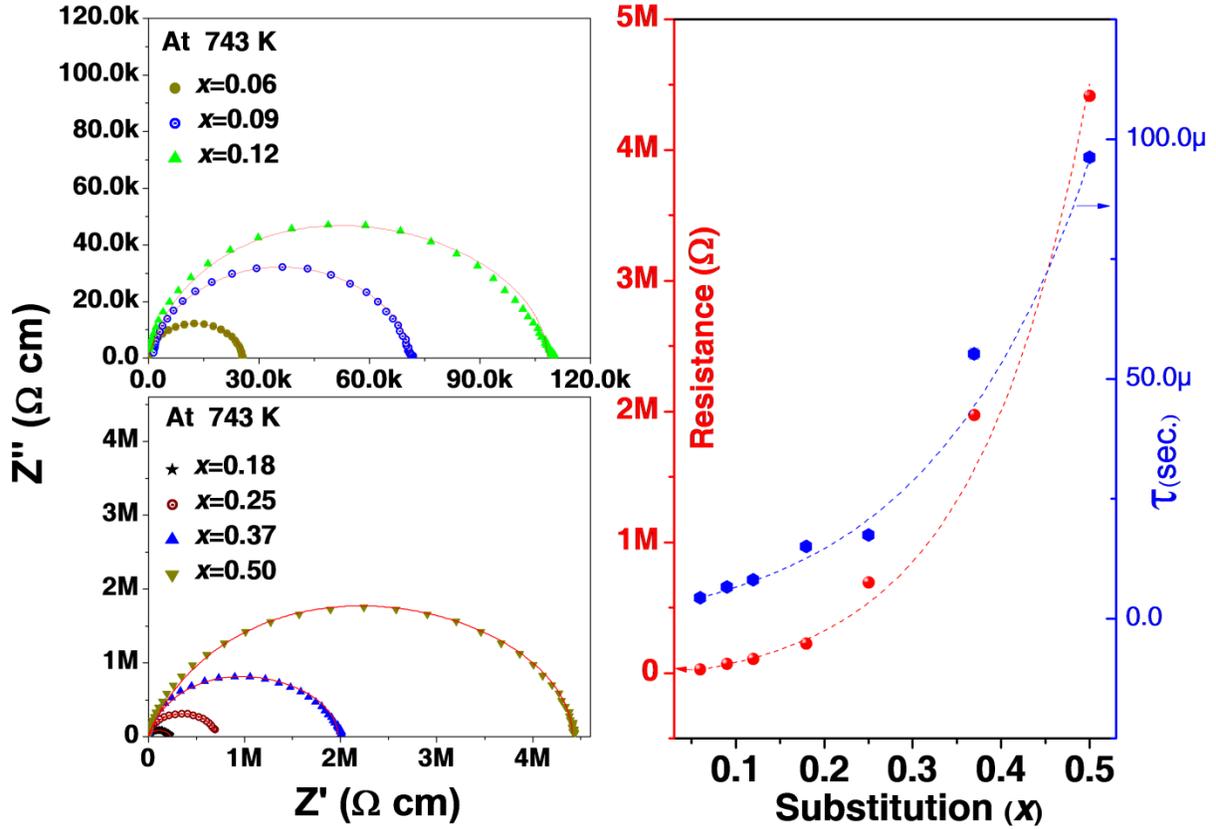

**Figure 5:** (Left) Nyquist plots of $Pb_{(1-x)}(K_{0.5}Sm_{0.5})_xTiO_3$ samples at 743K, (Right) resistance and time constant ($\tau$), extracted from fitting of Nyquist-plots by Cole-Cole equation for all samples.

The imaginary, $Z''$ (capacitive) and the real, $Z'$ (resistive) impedances are plotted in Figure 5. Capacitance due to grain boundary contribution is generally much higher than that of grain [37]. This may be due to grains having less resistance than grain boundary. In the Nyquist plot these effects are generally observed as three semicircles; the first smaller one belonging to grain, second grain boundary followed by the electrode contribution. Our results show only one semicircle. This contribution seems to be from the grain as the semicircle starts from origin (high frequency regime) and at high frequency we have observed a steady dielectric constant. Note that, it may possible due to less resistance at grain boundary, its effect suppresses and merge with grain effect. Also note that the resistance of the sample has increased with substitution figure 5 ( Right). The increasing resistance is an indication of better charge retention properties, enhancing the capability of being a polarizable material.

### 3.3 Ferroelectric properties

For $0.06 \leq x \leq 0.37$ the structure is elongated tetragonal type and a strained lattice is expected result in spontaneous polarization. It is to be noted that the tetragonality decreases with substitution until $x=0.5$ where the phase is cubic-like. This hints at reducing ferroelectricity with substitution. Also, the Raman modes broadens with increasing $x$ hinting at dispersed vibrational motion of the phonons. The $A_1$(1TO) mode is related to the vibration and distortion of the Ti-O bonds and thereby to ferroelectricity. The mode sustains for all samples although getting diminished with substitution. The dielectric properties show shifting of phase transition towards lower temperatures with increasing *x*, suggesting existence of ferroelectricity at room temperature.

[9]

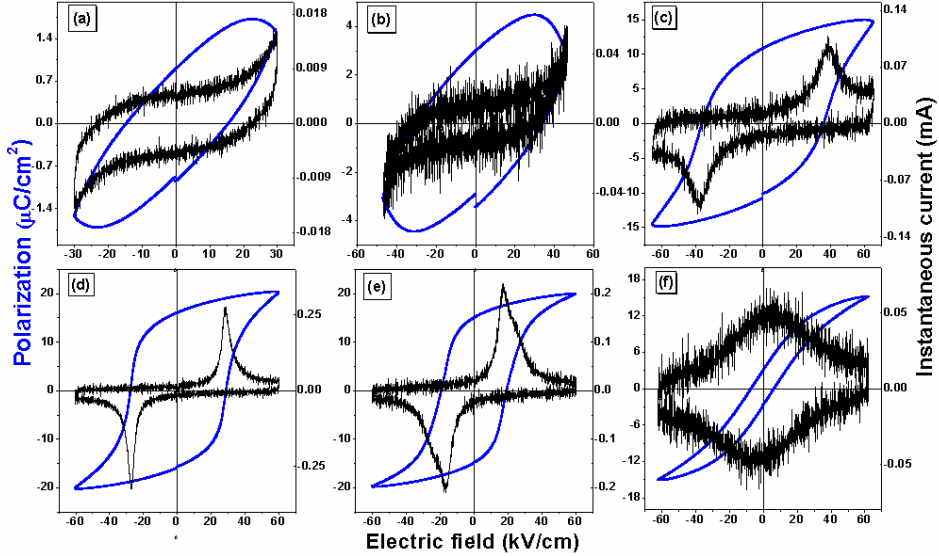

**Figure 6:** Polarization and instantaneous current versus electric field of $Pb_{(1-x)}(K_{0.5}Sm_{0.5})_xTiO_3$ samples, (a) *x*=0.06 (b) *x*=0.12 (c) *x*=0.18 (d) *x*=0.25 (e) *x*=0.37 (f) *x*=0.50.

The electric field dependent polarization (*P–E* hysteresis) and instantaneous current (*I–E* hysteresis) at 1 Hz frequency have been reported for all the compositions. For 0.06≤*x*≤0.12, *P-E* hysteresis loops [Figure 6] appears to be lossy dielectric type [39] rather than typical ferroelectric. Also, no signature of domain switching is observed in *I-E* data for *x*≤0.12 (Figure 6). The leaky, lossy nature may be due to comparatively lower resistances of the lesser modified samples (observed from impedance studies). For *x*>0.12, the leaky nature decreases and proper hysteresis *P-E* loops were observed. For *x*=0.5, it is expect a non-ferroelectric sample due to its cubic-like structure. However, we find a slim *P-E* loop with a weak *I-E* data. This hints at a nearly cubic structure for *x*=0.5. A detailed structural work needs to be done on this sample which is beyond the scope of this study. For a maximum applied field of 60 kV/cm the sample with *x*=0.25 shows highest apparent remnant polarization of 15.9 *μC/cm²* and a coercive field of 27.2 *kV/cm*. These values reduce with substitution to 2.97*μC/cm²* and 6.91*kV/cm* in *x*=0.5 composition.

Improvement of ferroelectric properties at room temperature in PKST samples results from increase of resistive and capacitive properties. In *PbTiO₃*, *Pb* and *O* vacancies are a common feature and results in non-stoichiometric samples mostly originating due to high volatility of *Pb* and subsequent *O*-vacancies due to missing *Pb* ions. By *K/Sm* substitution, *Pb* content is less, thereby evaporation of volatile matter is reduced. Therefore, *Pb* vacancies and resultant oxygen vacancies are reduced resulting in more stoichiometric samples. Although *K* is also volatile, the charge of the *K* ion is lesser than the *Pb* ion and hence *O* vacancies will be lesser. Vacancies play a very important role and enhance the conductivity of materials [40-45]. The defect structure can be represented by the Kröger–Vink notation [46]:

$$Pb_{Pb}^{\times} + O_o^{\times} \leftrightarrow PbO\ (\uparrow) + V_{Pb}'' + V_o^{\cdot\cdot} \qquad (6)$$

$$2K_K^{\times} + O_o^{\times} \leftrightarrow K_2O\ (\uparrow) + 2V_K' + V_o^{\cdot\cdot} \qquad (7)$$

Oxygen vacancies in perovskites can contribute to transport by migration from one site to another along the direction of applied high electric field increasing mobility and accumulate in the places with low free energy, such as domain walls and interfaces with electrodes. Accumulation of these oxygen vacancies at the domain boundary causes domain pinning. This restricts polarization switching [47].

## 4. Conclusion

($K_{0.5}Sm_{0.5}$) substituted $PbTiO_3$ ceramic samples were synthesized by sol-gel process. X-ray diffraction confirmed a tetragonal *P4mm* phase. Lattice parameter '*c*' decreases with increasing composition x, while 'a' is almost constant. Raman spectroscopy shows changes in intensity and energy of phonon modes related to tetragonal-cubic phase transformations and lattice strain with ferroelectric property. The lower mass of *K/Sm* compared to *Pb* may be a reason for such changes. However, in spite of the tetragonal- near cubic transformation the existence of Raman modes even for *x*=0.50 hints at the fact that either some local tetragonality is retained or some other phenomenon apart from a tetragonal structure generates strain which can be attributed to the remnant polarization for *x*=0.5 sample. High temperature dielectric study was clearly indicates a dielectric phase transition which moves towards lower temperature with increasing substitution. The diffuseness of the phase transition increases with substitution due to compositional disorder arising due to the random distribution of $K^+/Sm^{3+}$. As a result we find changes in ferroelectric property which is for lower substitutions not much of significance due to the leaky nature but with increasing resistance becomes more effective and promising as a ferroelectric. However the remnant polarization decreases with substitution probably due to reduction in tetragonality.

## Acknowledgement

One of the authors Arun K. Yadav is thankful to the university Grants Commission to award me fellowship (NFO-2015-17-OBC-UTT-28455).Principle investigator expresses sincere thanks to Indian Institute of Technology, Indore for funding the research and also using the Sophisticated Instrument Centre (SIC) for the research. The authors thank to Mr. Rituraj Sharma for providing the Raman facility in IISER Bhopal.

## References

1. H.S. Bhattia, S.T. Hussaina, F.A. Khanb and S. Hussainc, *Appl. Surf. Sci.*, 2016, **367**, 291-306.

[11]

2. H. Zhao, J. Miao, L. Zhang, Y. Rong, J. Chen, J. Deng, R. Yu, J. Cao, H. Wangd and X. Xing, *Dalton Trans.*, 2016, **45**, 1554-1559.
3. T.F. Zhang, X.G. Tang, Q.X. Liu, Y.P. Jiang, X.X. Huang and Q. F. Zhou, *J. Phys. D: Appl. Phys.*, 2016, **49**, 095302.
4. E.C. Paris, M.F.C. Gurgel, M.R. Joya, G.P. Casali, C.O. Paiva-Santos, T.M. Boschi, P.S. Pizani, J.A. Varela and E. Longo, *J. Phys. Chem. Solids*, 2010, **71**, 12-17.
5. G. Sághi-Szabó, R.E. Cohen and H. Krakauer, *Phys. Rev. Lett.*, 1998, **80**, 4321-4324.
6. N. Jaouen, A.C. Dhaussy, J.P. Itié, A. Rogalev, S. Marinel and Y. Joly, *Phys. Rev. B*, 2007, **75**, 224115-224120.
7. M. Adamiec, E. Talik and K. Wojcik, J. Alloys Compd., 2007, **442**, 222–224.
8. Q. Wei and J. Cong, *INTEGR FERROELECTR*, 2015, **167**, 35-40.
9. D.D. Shaha, P.K. Mehta, M.S. Desai, C.J. Panchal, *J. Alloys Compd.*, 2011, **509**, 1800-1808.
10. J.S. Forrester, J.S. Zobec, D. Phelan and E.H. Kisi, J. *Solid State Chem.*, 2004, **177**, 3553–3559.
11. I. Tomeno, J.A. Fernandez-Baca, K.J. Marty, K. Oka and Y. Tsunoda, *Phys. Rev. B,* 2012, **86**, 134306.
12. J. Chen, H. Shi, G. Liu, J. Cheng, S. Dong , *J. Alloys Compd.,* 2012, **537**, 280-285.
13. M. Alguero, P. Ramos, R. Jimenez, H. Amorın, E. Vila and A. Castro, Acta Mater., 2012, **60**, 1174-1183.
14. Y. Jia, H. Luo, S.W. Or, Y. Wang and H.L.W. Chan, *J. Appl. Phys.*, 2009, **105**, 124109.
15. X.P. Jiang, J.W. Fang, H.R. Zeng, B.J. Chu, G.R. Li, D.R. Chen and Q.R. Yin, *Mater. Lett.*, 2000, **44**, 219-222.
16. R.N.P. Choudhary and J. Mal, *Mater. Sci. Eng., B*, 2002, **90**, 1-6.
17. D.I. Woodward and I.M. Reaney, Acta Cryst., 2005, **B61**, 387-399.
18. I. Grinberg, M.R. Suchomel, P.K. davies and A.M. rappe, *J. Appl. Phys.*, 2005, **98**, 094111.
19. N. Raengthon, C.M. Cue and D.P. Cann, *J. Adv. Dielect.*, 2016, **6**, 1650002.
20. J.D. Freire and R.S. Katiyar, *Phys. Rev. B: Condens. Matter*, 1988, **37**, 2074-2085.
21. J. Frantti, V. Lantto, S. Nishio and M. Kakihana, *Phys. Rev. B,* 1999, **59**, 12-15.
22. G. Burns and B.A. Scott, *Phys. Rev. Lett.,* 1970, **25**, 167-170.
23. C. Sun, J. Wang, P. Hu, M.J. Kim and X. Xing, *Dalton Trans.*, 2010, **39**, 5183–5186.
24. V.R. Mastelaroa, P.P. Neves, S.R. de Lazaro, E. Longo, A. Michalowicz and J.A. Eiras, *J. Appl. Phys.*, 2016, **99**, 044104.
25. X. Chou, J. Zhai, H. jiang and X. Yao, J. Appl. Phys., 2007, **102**, 084106.
26. S. Kumar and K. B. R. Varma, J. Phys. D: Appl. Phys., 2009, **42**, 075405.
27. J. Cheng, S. Yu, J. Chen, Z. Meng, and L.E. Cross, *Appl. Phys. Lett.*, 2016, **89**, 122911.

28. http://ctcp.massey.ac.nz/Tablepol2014.pdf
29. A. Sani, M. Hanfland and D. Levy, *J. Solid State Chem.*, 2002, **167**, 446-452.
30. H. Sharma, J. Kreisel and P. Ghosez, *Phys. Rev. B,* 2014, **90**, 214102.

31. S. Bhaskar, S. B. Majumder and R.S. Katiyara, *Appl. Phys. Lett.*, 2012, **80**, 3997-3999.
32. J. Chen, Y. Qi, G. Shi, X. Yan, S. Yu and J. Cheng, *J. Appl. Phys.*, 2008, **104**, 064124.
33. A.A. Bokov, L.A. Shpak and I. P. Rayevsky, *J. Phys. Chem. Solids*, 1993, **54**, 495-498.
34. G. Deng, G. Li, A. Ding and Q. Yin, *Appl. Phys. Lett.*, 2005, **87**, 192905.
35. W. Ji, X. He, W. Cheng, P. Qiu, X. Zeng, B. Xia and D. Wang, *Ceram. Int.*, 2015, **41**, 1950-1956.
36. L. Liu, D. Shi, Y. Huang, S. Wu, X. Chen, L. Fang, C. Hu, Ferroelectrics, 2012, **432**, 65-72.
37. J.T.S. Irvine, D.C. Sinclair and A.R. West, Adv. Mater., 1990, 2,132-138.
38. S. Kumar and K.B.R. Varma, *Curr. Appl. Phys.*, 2011,**11**, 203-210.
39. J.F. Scott, *J. Phys.: Condens. Matter*, **20**, 2008, 20, 021001.
40. H. Beltrán, E. Cordoncillo, P. Escribano, D. C. Sinclair, and A. R. West, *J. Appl. Phys.*, 2005, **98**, 094102.
41. M. Li, J. Li, L.-Q. Chen, B.-L. Gu and W. Duan, *Phys. Rev. B*, 2015, **92**, 115435.
42. L. Liu, H. Fan, L. Fang, X. Chen, H. Dammak and M.P. Thi, Mater. Chem. Phys., 2009, **117**, 138-141.
43. 40 L. Liu, M. Wu, Y. Huang, Z. Yang, L. Fang, C. Hu, Mater. Chem. Phys., 2011, **126**, 769-772.
44. 41 L. Liu, Y. Huang, Y. Li, M. Wu, L. Fang, C. Hu, Y. Wang, Physica B, 2012, **407**, 136-139.
45. 42 L. Liu, Y. Huang, C. Su, L. Fang, M. Wu, C. Hu, H. Fan, Appl Phys A, 2011, **104**, 1047-1051.
46. F.A. Kröger, H.J. Vink, Solid State Physics, 1956, **3,** 307–435.
47. Y. Noguchi, I. Miwa, Y. Goshima, and M. Miyayamai, *Jpn. J. Appl. Phys.*, 2000, **39**, 1259-1262.